\documentclass[aip,tightenlines]{revtex4-1}
\usepackage{graphicx}
\usepackage{siunitx}
\usepackage[version=3]{mhchem}
\usepackage[italicdiff]{physics}

\begin{document}
\title{Thickness characterization of atomically-thin \ce{WSe2} on epitaxial graphene by low-energy electron reflectivity oscillations}
\author{Sergio C. de la Barrera}
\affiliation{Carnegie Mellon University, Department of Physics, Pittsburgh, PA 15213}
\author{Yu-Chuan Lin}
\affiliation{The Pennsylvania State University, Materials Science and Engineering, University~Park, PA 16802}
\author{Sarah M. Eichfeld}
\affiliation{The Pennsylvania State University, Materials Science and Engineering, University~Park, PA 16802}
\author{Joshua A. Robinson}
\affiliation{The Pennsylvania State University, Materials Science and Engineering, University~Park, PA 16802}
\author{Qin Gao}
\affiliation{Carnegie Mellon University, Department of Physics, Pittsburgh, PA 15213}
\author{Michael Widom}
\affiliation{Carnegie Mellon University, Department of Physics, Pittsburgh, PA 15213}
\author{Randall M. Feenstra}
\affiliation{Carnegie Mellon University, Department of Physics, Pittsburgh, PA 15213}

\begin{abstract}
In this work, low-energy electron microscopy is employed to probe structural as well as electronic information in few-layer \ce{WSe2} on epitaxial graphene on \ce{SiC}.
The emergence of unoccupied states in the \ce{WSe2}--graphene heterostructures are studied using spectroscopic low-energy electron reflectivity.
Reflectivity minima corresponding to specific \ce{WSe2} states that are localized between the monolayers of each vertical heterostructure are shown to reveal the number of layers for each point on the surface.
A theory for the origin of these states is developed and utilized to explain the experimentally observed featured in the \ce{WSe2} electron reflectivity.
This method allows for unambiguous counting of \ce{WSe2} layers, and furthermore may be applied to other 2D transition metal dichalcogenide materials.
\end{abstract}
\maketitle

\section{Introduction}
Low-energy electron microscopy (LEEM) is a powerful characterization tool for two-dimensional (2D) materials, since it provides both structural and electronic information, the latter dealing with unoccupied states above the surface vacuum level.
In such a system, a beam of electrons with energies between $0$ and \SI{20}{eV} is reflected from a sample surface at normal incidence.
The short penetration and escape depth of incident and reflected electrons with such low energy enables sensitivity to only the top-few atomic layers.
For these reasons, LEEM is highly suited to studies of 2D materials and 2D heterostructures.
There have been numerous LEEM studies of semimetallic graphene\cite{hibino2008microscopic,virojanadara2008homogeneous,sutter2009graphene,riedl2009quasi,locatelli2010corrugation,feenstra2013low87} and insulating hexagonal boron nitride,\cite{orofeo2013growth,gopalan2016formation} but the expanding class of 2D semiconductors remains to be investigated in detail.\cite{lin2014atomically,yeh2016direct,lin2016tuning}

Here, we prepare atomically-thin films of \ce{WSe2}, a semiconducting transition metal dichalcogenide (TMD), by metal-organic chemical vapor deposition (MOCVD) on epitaxial graphene on \ce{SiC}.
Epitaxial graphene (EG) provides an atomically-flat substrate for TMD growth and carries away excess charge during LEEM.
Low-energy electron diffraction (LEED) patterns taken from the surface indicate that the \ce{WSe2} crystals prepared by this method are crystalline and epitaxially aligned to the underlying graphene.
The preference for well-defined rotational alignment with graphene is promising for future electronic applications that require integration of 2D semiconducting and metallic components.

By measuring the reflected intensity of electrons as a function of effective beam energy, it is possible to extract spectroscopic information pertaining to electronic states at each point in the surface.
These spectra, called low-energy electron reflectivity (LEER), have been shown to allow unambiguous counting of the number of stacked monolayers of few-layer graphene and subsequent thickness mapping based on automated analysis methods.\cite{hibino2008microscopic,feenstra2013low87}
The layer-counting method relies on the presence of special states which are localized between the atomic layers of graphene, and on strong coupling between those states and the electrons involved in LEEM imaging.
Since \ce{WSe2} is a another layered material, it is a natural question to ask whether or not similar states exist between the quasi-2D layers of few-layer \ce{WSe2} and can be counted by analyzing electron reflectivity.
We show that by carefully considering features in the reflectivity of \ce{WSe2}, it is indeed possible to distinguish monolayer \ce{WSe2} on EG from regions with two layers or more.

\section{Methods}\label{sec:methods}
In this study, epitaxial graphene (EG) formed on 6H-\ce{SiC} is used as a template for synthesis of atomically-thin \ce{WSe2} crystals.
A \SI{1}{cm^2} piece of diced \ce{SiC} is etched in a \SI{10}{\percent} \ce{H2}/\ce{Ar} mixture at \SI{700}{Torr} and \SI{1500}{\celsius} for 30~minutes to remove surface damage caused by wafer polishing.
The \ce{SiC} is subsequently annealed in a pure \ce{Ar} environment at \SI{200}{Torr} and \SI{1620}{\celsius} for 10~minutes.\cite{lin2014direct}
During the entire process the \ce{SiC} substrates are inside a graphite crucible, which reduces the sublimation rate of \ce{Si} at high temperatures and hence improves the uniformity of graphene morphology.
\ce{WSe2} synthesis is carried out on EG substrates with conditions as previously reported by Eichfeld \emph{et al.},\cite{eichfeld2015highly} with the \ce{W} and \ce{Se} precursors in this growth being \ce{W(CO)6} and \ce{H2Se} respectively.

Following \ce{WSe2} growth, samples are transferred to an Elmitec LEEM III for characterization.
The principal mode of the LEEM directs a broad, monochromatic beam of electrons at the sample surface at normal incidence.
The elastically reflected electrons are filtered to allow only non-diffracted trajectories, and the remaining electrons are refocused into an image of the surface using a series of electron lenses.
Images are captured with a voltage bias applied between the sample surface and the electron gun, which determines the effective energy of the incident electrons.

Computations are performed using the Vienna Ab-Initio Simulation Package (VASP), employing the projector-augmented wave method and the Perdew-Burke-Ernzerhof generalized gradient approximation (PBE-GGA) to the exchange-correlation functional,\cite{kresse1993ab,kresse1996efficient,kresse1999ultrasoft,perdew1996generalized,*perdew1997generalized} with a plane-wave energy cutoff of \SI{500}{eV}.
Low-energy electron reflectivity (LEER) spectra of free-standing slabs of multilayer 2D materials are computed using a method described previously.\cite{feenstra2013low87,feenstra2013low130}
Inelastic effects are included in the computations,\cite{gao2015inelastic} employing an imaginary part of the potential, $V_i$.
Following the detailed analysis of Krasovksii and co-workers,\cite{krasovskii2002unoccupied,krasovskii2009very,flege2011self} in our prior work we employed the phenomenological expression $V_i = \SI{0.4}{eV}+0.06E$ where $E$ is the energy of a state relative to the vacuum level.\cite{gao2015inelastic}
These values for $V_i$ were found to give a reasonably good correspondence between experiment and theory, emphasizing experiments with energies of $0$--\SI{10}{eV}.
In the present work we are especially concerned with reflectivity behavior in the upper part of this range, near \SI{10}{eV} (and also including energies up to \SI{15}{eV}).
We find that use of the $V_i = \SI{0.4}{eV}+0.06E$ expression produces reflectivities that are too low (i.e. too much inelastic attenuation) near \SI{10}{eV}.
We therefore use a different expression, $V_i = \SI{0.4}{eV}+0.03E$, for all spectra computed here (i.e. the value of the slope parameter is reduced by a factor of $2$).
Comparing theoretical spectra obtained with these two expressions for $V_i$, we feel that this new expression might slightly underestimate inelastic effects near \SI{10}{eV} (and above) in typical 2D materials that we examine.
Nevertheless, this new expression provides a better means of examining such features in the theory since, again, attenuation near \SI{10}{eV} is significantly reduced.

\section{Experimental Results}\label{sec:exp-results}
\begin{figure*}
\includegraphics[width=6.750in]{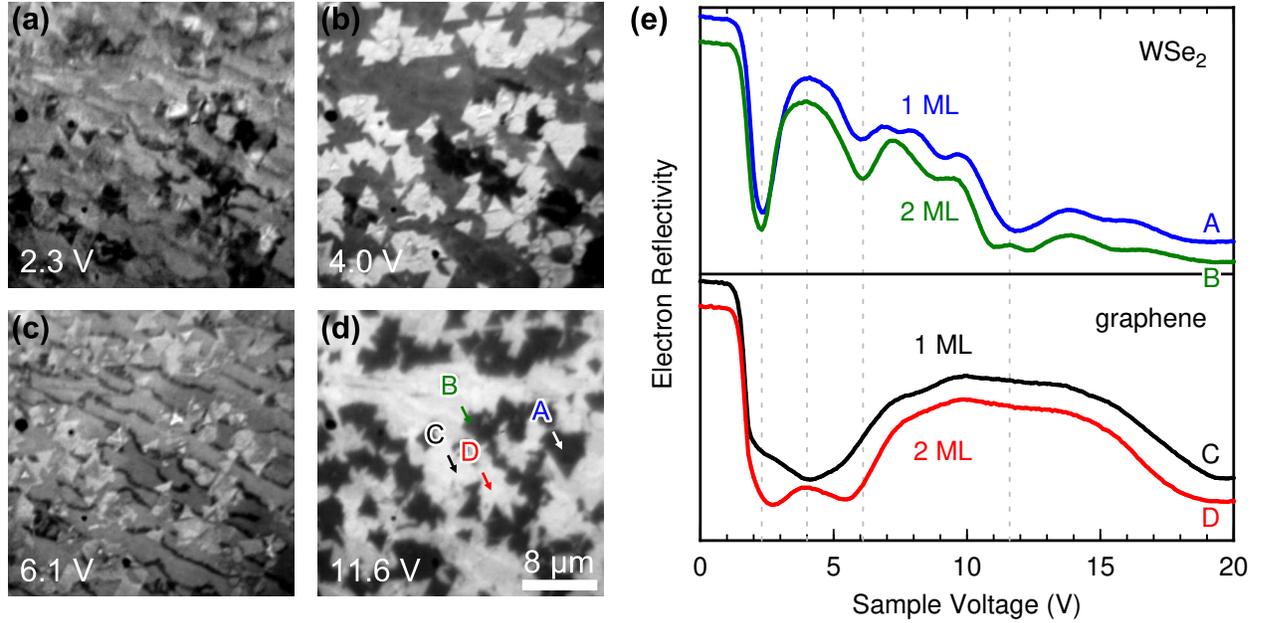}
\caption{\label{fig:images}
(a)--(d) LEEM images showing a single region of few-layer \ce{WSe2} crystals on epitaxial graphene on \ce{SiC} for a few sample bias voltages, as indicated.
(e) Reflected intensity of electrons extracted from the four labeled points in (d) as a function of sample voltage for two thicknesses of graphene and \ce{WSe2}.
Curves are shifted vertically for clarity and purposes of illustration.
Vertical dashed lines indicate the voltages used to capture the images in (a)--(d).
}
\end{figure*}

Figure~\ref{fig:images} shows LEEM images of the sample surface captured at a few sample voltages, showing the strong dependence of image contrast on sample bias.
This dependence can be quantified by recording the reflected intensity of electrons as a function of sample voltage for each pixel, in a series of images captured in a voltage sweep.
The resulting low-energy electron reflectivity (LEER) curves are extracted from the images for specific points or regions of interest to provide spectroscopic information about the surface.
For example, the reflectivity curves shown in Fig.~\ref{fig:images}(e) were extracted from the labeled points in panel \ref{fig:images}(d).
The relevant features in such spectra are reflectivity minima, which correspond to energies of electronic surface states that couple with incident electrons, causing transmission into the sample and thus reduced reflectivity at those energies.

The broad minimum in spectrum C of Fig.~\ref{fig:images}(e) near \SI{4.0}{V} is associated with a state that exists between monolayer graphene and the carbon-rich surface reconstruction of the \ce{SiC} below,\cite{hibino2008microscopic,feenstra2013low87} and therefore indicates the presence of monolayer graphene in that region of the image.
Curve D, which has two minima surrounding \SI{4.0}{V} and a local maximum in the middle, is similarly characteristic of bilayer graphene.
Curves A and B, however, originate from \ce{WSe2} regions, and yield a more complex set of reflectivity features with slight variations between the two curves.
The largest differences in these two curves are the shape of the minimum near \SI{6.1}{V} and the presence of a single- or double-minimum around \SI{11.6}{V}.

\begin{figure}
\includegraphics[width=3.375in]{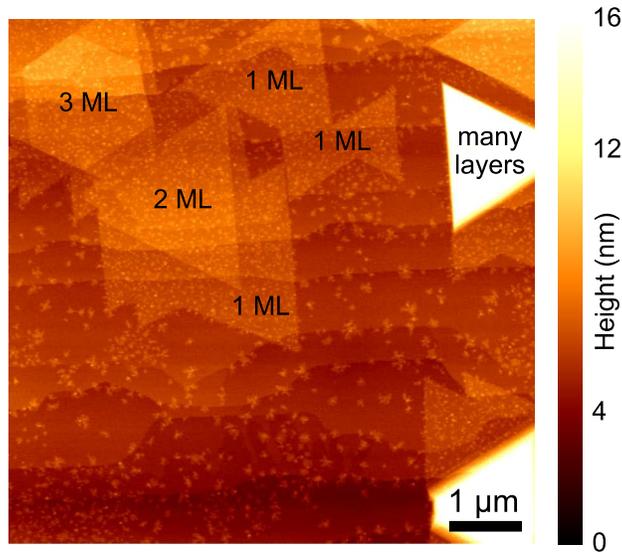}
\caption{\label{fig:afm}
Atomic force microscope image of surface height, showing monolayer ($1$~ML), bilayer ($2$~ML), and trilayer ($3$~ML) regions of \ce{WSe2} on the epitaxial graphene surface.
}
\end{figure}

\begin{figure*}
\includegraphics[width=6.750in]{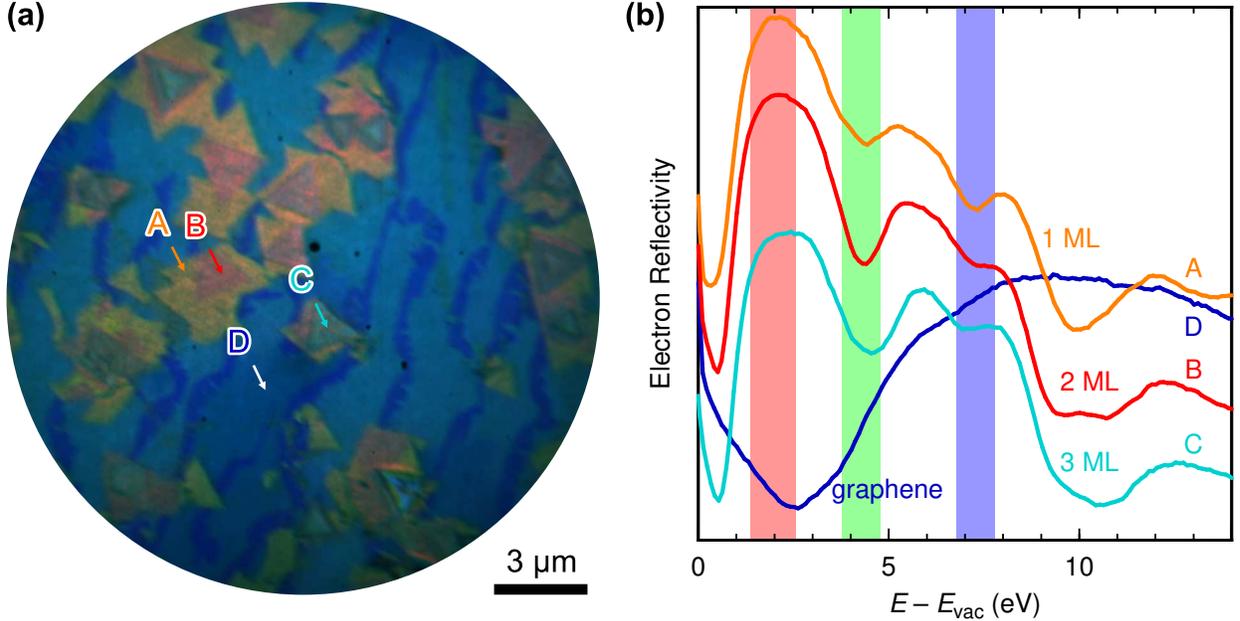}
\caption{\label{fig:cvd-refl}
(a) False-color spectroscopic image of MOCVD-grown \ce{WSe2} on epitaxial graphene, for the region shown in Fig.~\ref{fig:cvd-leem}, with colors assigned to the reflected intensity of electrons for specified energy windows.
(b) Reflected intensity of electrons from labeled locations in (a).
Curves are shifted vertically for clarity and plotted versus energy, rather than sample voltage, for comparison with theory.
Colored energy ranges indicate those used to generate the spectroscopic image.
}
\end{figure*}

Atomic force microscope (AFM) scans of the surface reveal that the majority of the \ce{WSe2} crystals are monolayer ($1$~ML) and bilayer ($2$~ML), with a few instances of thicker island growth (Fig.~\ref{fig:afm}).
The height change between the top of a monolayer crystal and the EG surface is approximately \SI{0.65}{nm}, similar to other samples prepared by the same method.\cite{eichfeld2015highly,lin2015atomically}
Electron reflectivity from one of these monolayer \ce{WSe2} crystals is shown in curve A of Fig.~\ref{fig:cvd-refl}(b), with a local minimum at \SI{10}{eV}.
We ascribe the occurrence of this minimum to a specific state which exists in monolayer \ce{WSe2}, and will be discussed in Section~\ref{sec:theory-results}.
Bilayer \ce{WSe2} triangles are also observed in LEEM as well as AFM.
The reflectivity from one of these triangles, shown in curve B of Fig.~\ref{fig:cvd-refl}(b), exhibits two reflectivity minima surrounding a local maximum at \SI{10}{eV}.
In this case the two minima can be understood to result from a combination of two nearly-degenerate monolayer-\ce{WSe2} states, and thus this double-minimum is a signature of bilayer \ce{WSe2}.

\begin{figure}
\includegraphics[width=3.375in]{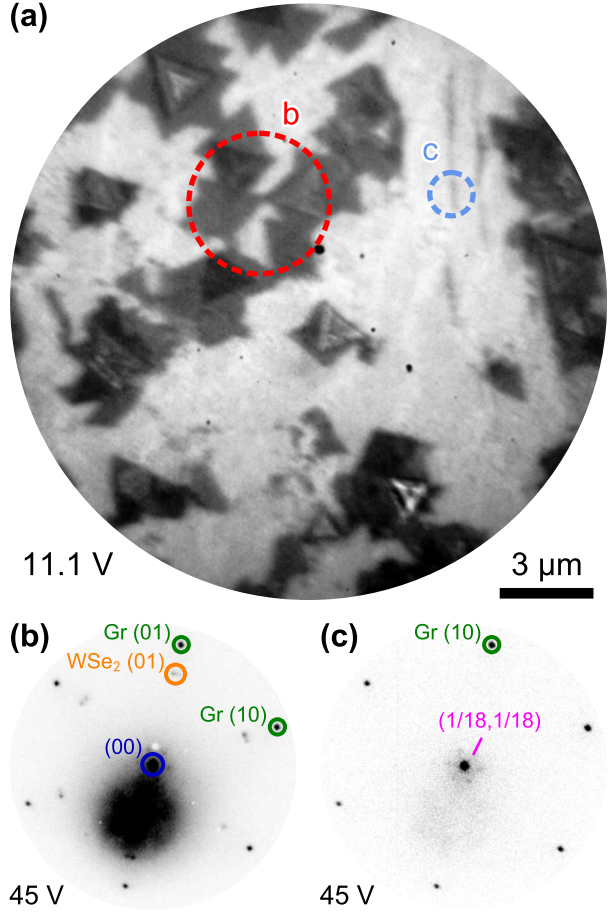}
\caption{\label{fig:cvd-leem}
(a) LEEM image of \ce{WSe2}--EG--\ce{SiC}, showing \SI{2}{\micro\metre} triangular \ce{WSe2} islands on a bright background of few-layer epitaxial graphene.
(b) Selected-area diffraction from the circular region labeled ``b'' in the LEEM image shows six dark, outer spots from the graphene lattice, with six additional groups of spots associated with the \ce{WSe2} islands at a smaller wavevector.
Surrounding the non-diffracted $(00)$ spot, there are six satellite spots associated with the $6\sqrt{3}\times6\sqrt{3}$--R\SI{30}{\degree} reconstruction of the \ce{SiC}.
(c) Diffraction from the bare graphene region labeled ``c'' in the LEEM image shows only the six outer diffraction spots and the $6\sqrt{3}$ structure also found in (b), labeled by $(\flatfrac{1}{18},\flatfrac{1}{18})$.
}
\end{figure}

To classify the crystals within the imaged region in Fig.~\ref{fig:cvd-leem}(a), we create a colorized map based on the relevant reflectivity features.
Colors are assigned based on the total reflectivity of specific energy windows for each point on the surface, and the result is a false-color spectroscopic image, weighted by the spectral components within each energy window, as in Fig.~\ref{fig:cvd-refl}(a).
From this spectroscopic image, we clearly see the few-layer graphene areas, which primarily have states within the band gap region of the \ce{WSe2} spectrum (between $1.5$ and \SI{3.5}{eV}, with high reflectivity) and appear blue due to the assignment of red and green channels to energies in this regime.
Two \ce{WSe2} reflectivity minima near $4$ and \SI{7}{eV}, which evolve with the number of layers, are assigned to green and blue channels, respectively, causing color variations in the map based on the number of layers.
For example, in this color scheme, monolayer \ce{WSe2} appears yellow-hued, while bilayers appear rose-hued, and trilayers appear turquoise (for a few small triangles in the center of pyramidal structures).
The map generated by this colorization scheme is further evidence of the reproducibility of reflectivity analysis for determining \ce{WSe2} thickness.

In another mode of LEEM operation, diffraction patterns are acquired, allowing direct analysis of the surface structure.
We insert a small aperture to reduce the illuminated area of the surface and collect a diffraction pattern for the local region, so-called selected area diffraction or \si{\micro}LEED.
Diffraction patterns from the encircled regions in Fig.~\ref{fig:cvd-leem}(a) show distinct 6-fold diffraction spots from the graphene (larger wavevector) and \ce{WSe2} (smaller wavevector), with six additional satellite spots surrounding the central, specular $(00)$-spot, originating from the $6\sqrt{3}\times6\sqrt{3}$--R\SI{30}{\degree} surface reconstruction, also known as the buffer layer of EG--\ce{SiC}.\cite{virojanadara2008homogeneous}
The \ce{WSe2} spots form small groups azimuthally-centered on the diffraction pattern of the underlying graphene.
From the angular spread of these points, we find that the \ce{WSe2} preferentially forms rotationally aligned with the graphene lattice, within \SI{\pm2.3}{\degree}, for the given growth conditions.
Interestingly, the macroscopic alignment of the triangular crystal edges seen in the LEEM images are primarily oriented within \SI{60}{\degree} of one another.
This suggests that a specific edge termination is preferred by this growth method, however, from LEEM it is not clear which type.

\section{Theoretical Results}\label{sec:theory-results}
As first discussed by Hibino \emph{et al.}\cite{hibino2008microscopic,hibino2008thickness} and extensively analyzed in our prior work,\cite{feenstra2013low87,feenstra2013low130,gao2015inelastic,srivastava2013low} the occurrence of minima in low-energy electron reflectivity spectra is associated with interlayer states that occur between the 2D planes of van der Waals (vdW) bonded materials.
Such interlayer states arise from the image-potential states that exist on either side of a single 2D layer,\cite{posternak1983prediction,silkin2009image} i.e. a monolayer (ML) of carbon for the case of graphene or ML-\ce{WSe2} for the case of bulk \ce{WSe2}.
When 2D MLs are brought together to form a vdW-bonded bulk material, the image-potential states of the respective layers combine to form a band of interlayer states.\cite{posternak1983prediction}
The image-potential states themselves have energies some $10$'s of \si{meV} below the vacuum level, but when they combine to form the interlayer states then those states end up with energies typically in the range of $0$--\SI{8}{eV} above the vacuum level, at least for the case of graphene.
\footnote{
The interlayer band that we are discussing here, occurring in the $0$--\SI{8}{eV} range, is actually the lowest band of a pair of two bands.
The upper band has energy of $14$--\SI{22}{eV}, at least for the case of graphene.\cite{feenstra2013low130}
The combinations of image-potential states that form the bands are symmetric (antisymmetric) for the lower (upper) band, relative to a location midway between the sheets of 2D material.
}
As discussed in prior work, the interlayer states are free-electron like,\cite{posternak1983prediction} in the sense that in the spaces between the 2D sheets (the interlayer spaces), these states have character similar to that of plane wave with wavevector magnitude of $\kappa_0 = \sqrt{2m(E-E_\mathrm{vac})}/\hbar$ where $E-E_\mathrm{vac}$ is the energy of the state relative to the vacuum level.
The wavefunctions of the interlayer states tend to be concentrated in the interlayer spaces; they have a local maximum at a location midway between neighboring 2D planes.

Given a band structure of a vdW-bonded bulk material, we analyze it to determine the amount of plane-wave character within the interlayer space that each state exhibits.
With the $z$-direction being along the $c$-axis of the material, it is only necessary to consider states with wavevector components $(k_x, k_y) = (0, 0)$ and $k_z \equiv k$.
We define an overlap between a wave function of the material and a plane wave according to:
\begin{subequations}\label{eq:sigma}
\begin{align}
\sigma_\pm &= \frac{\sqrt{Ac}}{z_2 - z_1} \int_{z_1}^{z_2} \phi_{\nu,\pm k}^{0,0}(z)\exp(i\kappa_0 z) \dd{z},\,\text{and} \\
\sigma &\equiv (\abs{\sigma_+}^2 + \abs{\sigma_-}^2)^{1/2},
\end{align}
\end{subequations}
where $A$ is the area of the lateral unit cell of the material and $c$ is the $c$-axis periodicity, $z_1$ and $z_2$ define the interlayer space over which the overlap is computed, and $\phi_{\nu,\pm k}^{0,0}(z)$ is the $(G_x, G_y) = (0, 0)$ Fourier coefficient of the wave function (equal to the wave function averaged over the lateral unit cell).
We note that this form is the same as the one we previously introduced in connection with our low-energy reflectivity analysis, although in that prior analysis it was evaluated for the case of far-separated 2D layers in a periodic supercell,\footnote{
An additional distinction between the form introduced in Eq.~\ref{eq:sigma} and that used previously in Ref.~\onlinecite{feenstra2013low87} is that the former refers to states with $\pm k_z$ whereas the latter referred to even and odd states formed by linear combinations of the $\pm k_z$ states.
However, the resulting values for $\sigma$ are identical for both cases.
} whereas in the present case it is evaluated between $2$~ML of a bulk material.
All of the evaluations of $\sigma$ presented below are performed by computing the overlap over a $2$-\si{\angstrom}-wide space centered at the midpoint of the interlayer space, with $z_2 - z_1 = \SI{2}{\angstrom}$.

\begin{figure}
\includegraphics[width=3.125in]{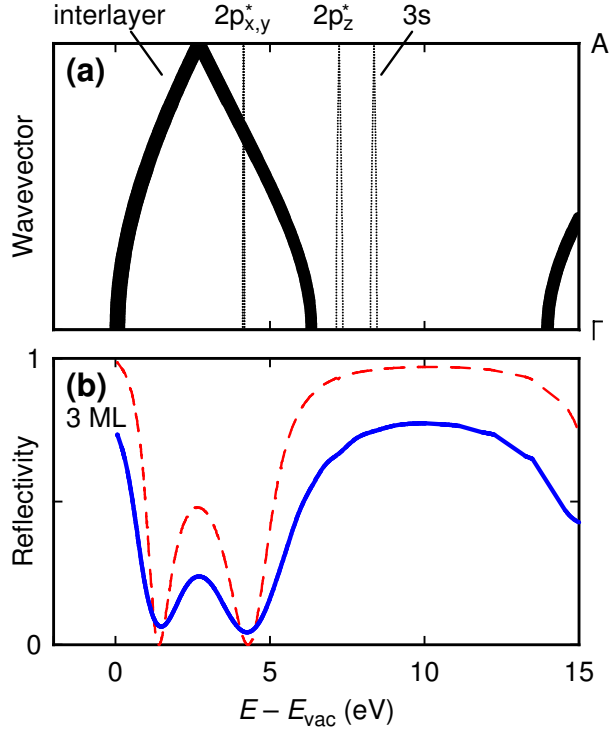}
\caption{\label{fig:comp-graphene}
(a) Band structure of graphite, with wavevector varying from $\Gamma$ to $\mathrm{A}$.
Symbol sizes, beyond a minimum size, are proportional to the value of $\sigma$ (Eq.~\ref{eq:sigma}) for each state.
(b) Computed LEER spectra of $3$~ML free-standing graphene, with (blue solid line) and without (red dashed line) inelastic effects.
Energies are relative to the vacuum level of the $3$-ML slab.
}
\end{figure}

Before examining the band structure for the material of interest, \ce{WSe2}, it is instructive to first review the situation for simpler materials such as graphite and hexagonal boron nitride (h-\ce{BN}).
Figure~\ref{fig:comp-graphene}(a) shows the band structure of graphite, for $(k_x, k_y) = (0, 0)$.
We use symbol sizes for the plotting which, for each state, are given by some minimum symbol size plus an amount that is proportional to the computed value of $\sigma$ for that state.
Hence, bands that have significant plane-wave character (i.e. significant interlayer character) are revealed by the relatively large symbol sizes.
As is well known from prior work,\cite{feenstra2013low87,hibino2008microscopic} in graphite there is only a single band with interlayer character, the one labeled ``interlayer'' at the top of Fig.~\ref{fig:comp-graphene}.
Importantly, this interlayer band has its origin not in terms of any atomic orbitals in the material, but rather, it arises from plane waves existing in the interlayer spaces as already discussed above.
All the other bands that are seen in Fig.~\ref{fig:comp-graphene}(a), however, \emph{can} be related to specific combinations of atomic orbitals, as labeled at the top of the figure. 

The situation for graphite is especially simple since there is zero coupling (zero overlap) between the interlayer band and the overlapping and/or nearby bands.
Specifically, we consider the bands labeled $2p_{x,y}^*$, $2p_z^*$, and $3s$ in Fig.~\ref{fig:comp-graphene}(a).
These labels are meant to be approximate ones, indicative of the character of the states in the bands.
This character is readily apparent from several types of analysis; examination of the spherical symmetry of the states relative to atomic locations, tight-binding modeling of the bands and comparison to first-principles results, examination of the dependence of the bands on interlayer separation, and individual inspection of specific wavefunctions of the states.\cite{mende2015growth}
We find that all of the states of these three bands are orthogonal to the states in the interlayer band.
This orthogonality arises for the states of the $2p_{x,y}^*$ band due to its composition in terms of in-plane $p$-orbitals, whereas it arises for the $2p_z^*$ and $3s$ bands because the wave functions of states in those bands have opposite sign on neighboring \ce{C} atoms of the graphene lattice.

Figure~\ref{fig:comp-graphene}(b) shows the low-energy electron reflectivity (LEER) spectrum that arises from free-standing multilayer graphene containing $3$ graphene layers, computed without and with inelastic effects.
As is well known from prior work, one reflectivity minimum occurs for every interlayer space in the structure.
For example, for $3$ graphene layers there are $2$ interlayer spaces and hence $2$ reflectivity minima.\cite{feenstra2013low87}
The theoretical spectrum including inelastic effects shown in Fig.~\ref{fig:comp-graphene}(b) is in good agreement with experiment.\cite{gao2015inelastic,hibino2008microscopic}
Importantly, since there is no overlap between the states of the interlayer band and those of overlapping and/or nearby bands, those bands make no contribution to the resulting LEER spectra. 

\begin{figure}
\includegraphics[width=3.125in]{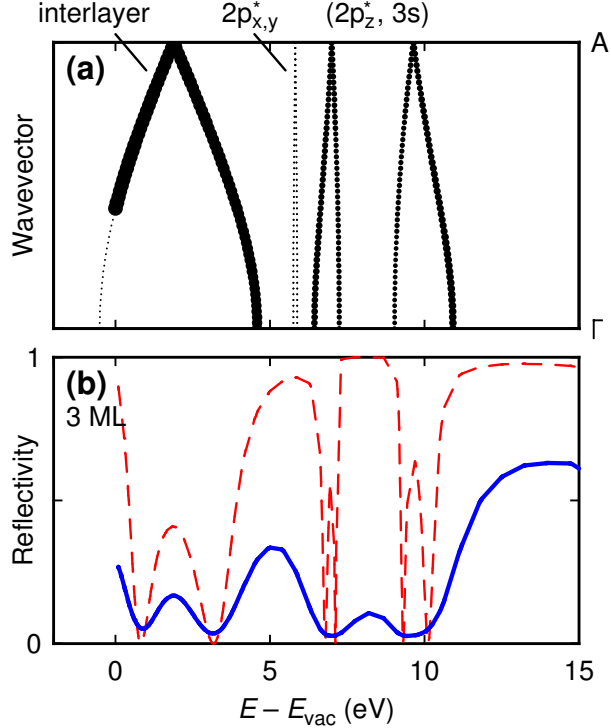}
\caption{\label{fig:comp-hBN}
(a) Band structure of bulk h-\ce{BN}, with wavevector varying from $\Gamma$ to $\mathrm{A}$.
Symbol sizes, beyond a minimum size, are proportional to the value of $\sigma$ (Eq.~\ref{eq:sigma}) for each state.
(b) Computed LEER spectra of $3$~ML of free-standing h-\ce{BN}, with (blue solid line) and without (red dashed line) inelastic effects.
Energies are relative to the vacuum level of the $3$-ML slab.
}
\end{figure}

In Fig.~\ref{fig:comp-hBN} we display results for h-\ce{BN}.
Figure~\ref{fig:comp-hBN}(a) shows the bulk h-\ce{BN} band structure, again with symbol sizes computed in accordance with the $\sigma$ values.
The inequivalence between the \ce{B} and \ce{N} atoms of h-\ce{BN} produces large changes to the band structure compared to that of graphene, but nevertheless, a single interlayer band together with a few nearby bands can be identified in Fig.~\ref{fig:comp-hBN}(a).
One of these nearby bands has $2p_{x,y}^*$ character; as for graphene, the states of this band are orthogonal to states of the interlayer band.
However, in contrast to the situation for graphene, the other two nearby bands, which for h-\ce{BN} have mixed $2p_z^*$ and $3s$ character, are not orthogonal to the interlayer band.
This difference occurs simply due to the inequivalence of \ce{B} and \ce{N} atoms, which destroys the precise orthogonality described above for graphite.
Hence, these two nearby bands acquire some degree of plane-wave (interlayer) character.

Resulting LEER spectra for 3~MLs of free-standing h-\ce{BN}, with and without inelastic effects, are displayed in Fig.~\ref{fig:comp-hBN}(b).
In the absence of inelastic effects, the coupling of the interlayer character with two of the nearby bands leads to reflectivity minima associated with each of the bands.
All of the three bands with interlayer character in Fig.~\ref{fig:comp-hBN}(b) display two reflectivity minima each, arising from the two interlayer spaces.
However, when inelastic effects are included, a large amount of broadening occurs in the spectra, particularly for the two bands with mixed $2p_z^*$ and $3s$ character.
The reflectivity maximum that occurs at \SI{8.2}{eV} between these two bands for the computation neglecting inelastic effects is greatly diminished in size, to become a weak, local maximum which separates the two minima (at 7.0 and \SI{9.5}{eV}) of this band.
No discrete thickness oscillations are observed in connection with these minima; the oscillations found in the absence of inelastic effects are eliminated when the inelastic effects are included.
Experimentally, a broad reflectivity minimum centered at about \SI{8.2}{eV} above the vacuum level has indeed been observed in h-\ce{BN} LEER spectra,\cite{orofeo2013growth,gopalan2016formation} and two minima (or a minimum and a shoulder) are seen within that broad minimum.
As mentioned in Section~\ref{sec:methods}, for the computation of Fig.~\ref{fig:comp-hBN}(b) we are employing values for the energy-dependent imaginary part of the potential (which governs inelastic effects) which are somewhat reduced from our typical values, in order to emphasize these features in the $7$--\SI{11}{eV} range (which are especially relevant for the \ce{WSe2} spectra).

\begin{figure}
\includegraphics[width=3.125in]{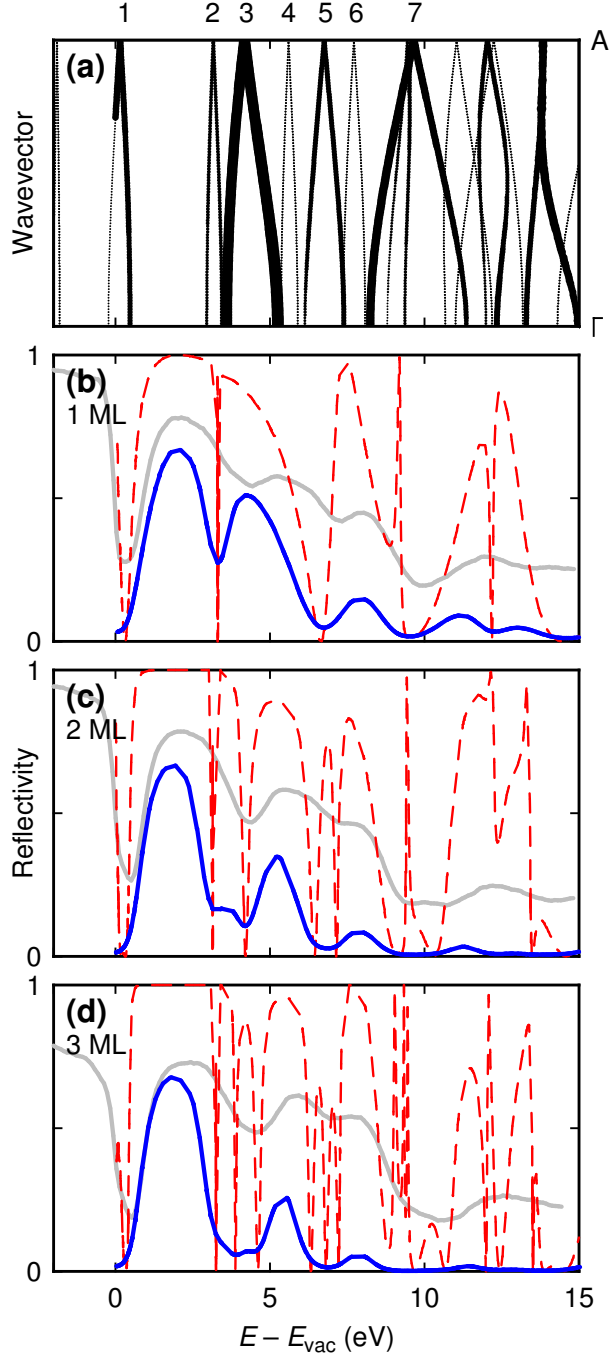}
\caption{\label{fig:comp-WSe2}
(a) Band structure of bulk \ce{WSe2}, with wavevector varying from $\Gamma$ to $\mathrm{A}$.
Symbol sizes, beyond a minimum size, are proportional to the value of $\sigma$ (Eq.~\ref{eq:sigma}) for each state.
(b)--(d) Computed LEER spectra of $1$, $2$, and $3$ ML of free-standing \ce{WSe2} as indicated, with (blue solid lines) and without (red dashed lines) inelastic effects.
Experimental curves (gray solid lines) from Fig.~\ref{fig:cvd-refl} are superimposed for comparison.
Energies are relative to the vacuum level of the respective slabs.
}
\end{figure}

Figure~\ref{fig:comp-WSe2} displays the bulk bands for \ce{WSe2}.
There are many more bands than for graphene or h-\ce{BN}, arising from the multiplicity of $s$, $p$, and $d$ states of the \ce{W} and \ce{Se} atoms.
Low-lying bands of interest in Fig.~\ref{fig:comp-WSe2}(a) are numbered $1$--$7$ (with band $7$ being the relatively wide band with significant plane-wave character centered at \SI{10}{eV}).
From a decomposition of the states into their $s$, $p_{x,y}$, $p_z$, $d_{z^2}$, $d_{xz,yz}$, and $d_{xy,x^2-y^2}$ character (not shown), we find that bands $4$ and $6$, each of which is doubly degenerate, have purely $d_{xz,yz}$ character, with nodal planes parallel to the $xz$ and $yz$ planes.
Hence, these bands have no plane-wave character, and they make no contribution to the reflectivity.
Of the remaining bands, band $3$ is seen to have the most plane-wave character, bands $1$ and $7$ have substantial plane-wave character, and bands $2$ and $5$ have a small amount of plane-wave character. 

Reflectivity for free-standing slabs of $1$, $2$, and $3$ MLs of \ce{WSe2} are shown in Figs.~\ref{fig:comp-WSe2}(b)--\ref{fig:comp-WSe2}(d), respectively.
The spectra that do not include inelastic effects reveal thickness oscillations for most of the bands, with the number of minima given by either the number of layers ($n$) or the number of interlayer spaces ($n-1$), depending on the particular band.
However, with inelastic effects included all of these oscillations disappear, and the respective minima associated with each band appear just as a single, broad minimum.
These broad minima occur at approximately the same energies (relative to the vacuum level) as the features observed in the experimental spectra of Section~\ref{sec:exp-results}.
For comparison, these experimental curves are reproduced in Fig.~\ref{fig:comp-WSe2} as well.

Concerning the small reflectivity features discussed in Section~\ref{sec:exp-results} at $4$ and \SI{10}{eV} which we associate with differing thicknesses of the \ce{WSe2}, these are more difficult to discern in the theoretical spectra.
However, comparing the $1$~ML and $2$~ML spectra, we see a significant difference in their behavior near \SI{10}{eV}; the former shows a single, distinct minimum at \SI{9.7}{eV}, whereas the latter displays a broad minimum extending over about $9.0$--\SI{10.5}{eV} (with two minima in the elastic-only computation seen at either end of this range).
For the case of $3$~ML of \ce{WSe2}, an even broader minimum near \SI{10}{eV} is seen.
Of course, an important distinction between the theoretical spectra of Fig.~\ref{fig:comp-WSe2} and the experimental spectra of Section~\ref{sec:exp-results} is that the former are for free-standing \ce{WSe2} MLs, whereas the latter are for \ce{WSe2} on top of an epitaxial graphene substrate.
This difference is further discussed in the following section.

\section{Discussion}
Computation of reflectivity spectra for \ce{WSe2} on few-layer graphene is quite complex due to the poor epitaxial fit of the materials and the large size of the supercell required.
Nevertheless, predictions for the evolution of reflectivity minima for free-standing slabs of $1$-, $2$-, and $3$-ML \ce{WSe2} appear to be sufficient for interpretation of the experimentally-measured reflectivity from \ce{WSe2}--EG--\ce{SiC}, despite neglecting the effect of the substrate.
In comparing the measured results from Section~\ref{sec:exp-results} to the computed reflectivity in Section~\ref{sec:theory-results}, it is important to note that the experimental curves are measured versus sample voltage $V_S$, and not energy above the vacuum level $E-E_\mathrm{vac}$ directly.
Due to the work function difference $\Delta W$ between the electron gun filament of the LEEM and the \ce{WSe2} on the sample surface, the experimental curves are shifted approximately \SI{2.2}{V} (depending on location) toward higher voltage.
Using a quantitative method for determining the local vacuum level outlined in Ref.~\onlinecite{gopalan2016formation}, the experimental reflectivity curves are shifted by $\Delta W$ in order to plot the spectra versus $E-E_\mathrm{vac} = eV_S - \Delta W + \sigma_c$, including a small energy shift $\sigma_c \approx \SI{0.1}{eV}$ to account for the peak energy of thermionic emission from the gun cathode.
With this method in place, it is possible to plot the experimental reflectivity curves together with the computed ones in Fig.~\ref{fig:comp-WSe2}.

It is a known result that high-energy bands computed with PBE-GGA (as discussed in Section~\ref{sec:methods}) are generally lower energy than real bands.
As such, the subsequent computed reflectivity curves are typically shifted $0.5$ to \SI{1}{eV} lower along the energy axis compared to experiment.\cite{mende2015growth}
With this in consideration, we conclude that there is reasonable agreement between the computed and experimental minima near $0$, $7$, and \SI{10}{eV}.

Critically, the minimum near \SI{10}{eV} in the $1$~ML computed reflectivity curve shown in Fig.~\ref{fig:comp-WSe2}(b) evolves into a broad, flat minimum in the $2$~ML case, as in \ref{fig:comp-WSe2}(c).
The flat minimum occurs in the computed reflectivity due to the combined effect of two nearby states, one of which has lower energy and produces a deeper minimum in the $2$-ML case than in the $1$-ML case.
The elastic-only computed curves show this behavior most clearly, although the overall effect becomes complicated for more than $2$~ML.
A similar flattening of the minimum near \SI{10}{eV} is clearly observed in the $1$- and $2$-ML experimental curves (gray solid lines in Fig.~\ref{fig:comp-WSe2}(b) and \ref{fig:comp-WSe2}(c)), although in the measured curves there are two distinct minima, whereas our best fit shows no clear oscillations using the inelastic model implemented here.
In any case, beyond $2$~ML it may be difficult to resolve additional minima in measured reflectivity due to inelastic effects.

The states which form band~$3$ have strong interlayer character and subsequently vary as the number of interlayer spaces, $n-1$.
In addition, states from nearby band~$2$ couple and broaden the resulting reflectivity minimum such that for $1$~ML of \ce{WSe2}, there is a narrow minimum near \SI{3.3}{eV}, whereas for $2$~ML the minimum is deeper and shifted to higher energy.
This effect is also observed in the experimental reflectivity outlined in Section~\ref{sec:exp-results} and therefore provides another signature for discriminating between $1$- and $2$-ML \ce{WSe2}.
For a greater number of layers, the computed minimum near \SI{4}{eV} is expected to broaden and deepen further, but will not develop countable oscillations like those near \SI{10}{eV}.
It is the wide dispersion of band~$7$ that allows the states in the few-layer limit to be resolved, as was the case for the interlayer bands in graphene and h-\ce{BN}.
Thus, for bands with small dispersion the variation with number of layers is predicted to be less pronounced.

Finally, although the computations considered here do not include the graphene or \ce{SiC} below the \ce{WSe2} layers, it is reasonable to posit that interactions between the \ce{WSe2} and graphene might have an effect on the reflectivity.
In particular, minima associated with interlayer states in few-layer graphene occupy an energy window from $0$--\SI{7}{eV}, as in Fig.~\ref{fig:comp-graphene}.
The band gap in the \ce{WSe2} spectrum between bands $1$ and $2$ reflects most electrons with energy in that range, and therefore prevents coupling to graphene interlayer states below the \ce{WSe2}, however, there may still be coupling between the upper \ce{WSe2} band gap edge and \SI{7}{eV}.
Whether or not evidence of this can be observed remains an open question.

\section{Conclusions}
We have shown that low-energy electron reflectivity measurements of \ce{WSe2}--EG--\ce{SiC} yield distinct spectroscopic signatures for \ce{WSe2} and graphene regions.
By correlating the observed LEEM images with AFM scans of the surface, we have identified monolayer and bilayer crystals of \ce{WSe2} and labeled the reflectivity accordingly.
Using a first-principles method of calculating electron reflectivity curves from free-standing slabs of few-layer \ce{WSe2}, we have assigned the observed features in $1$- and $2$-ML-\ce{WSe2} reflectivity to specific states with strong plane-wave character.
We argued that enumeration of these states provides a clear evolution of reflectivity minima as layer number increases, and that this evolution allows discrimination between $1$- and $2$-ML-\ce{WSe2} from the reflectivity alone.
Furthermore, by numerically analyzing the spectral features from a LEEM imaging dataset it is possible to generate a colorized map of \ce{WSe2} layer thickness with high fidelity across the image.
This method paves a path forward for quickly determining few-layer \ce{WSe2} film thickness with atomic resolution, and may be applicable to other TMD materials as well.
The results and analyses presented here provide critical insight for future studies of layered heterostructures including \ce{WSe2} and graphene, as well as LEEM studies of other 2D materials.

\section{Acknowledgements}
We are grateful to P. Mende and S. Satpathy for useful discussions.
This work was supported in part by the Center for Low Energy Systems Technology (LEAST), one of six centers of STARnet, a Semiconductor Research Corporation program sponsored by MARCO and DARPA.

\bibliography{/research/ref/abbr/shortjournals,/research/ref/all/refs}
\end{document}